\documentclass[
aps,%
12pt,%
final,%
notitlepage,%
oneside,%
onecolumn,%
nobibnotes,%
nofootinbib,%
superscriptaddress,%
noshowpacs,%
centertags]%
{revtex4}

\begin{document}
\selectlanguage{english}

\title{Probing cosmic plasma with giant radio pulses}

\author{\firstname{V.~I.} \surname{Kondratiev}}
\email{vlad@asc.rssi.ru}%
\affiliation{%
\mbox{Astro Space Center of the Lebedev Physical Institute, Profsoyuznaya 84/32,} Moscow, 117997 Russia
}%
\affiliation{%
\mbox{Department of Physics, West Virginia University, PO Box 6315, Morgantown, WV 26505, USA}
}%
\affiliation{%
\mbox{York University, Department of Physics and Astronomy, 4700 Keele Street, Toronto,} Ontario M3J 1P3 Canada
}%

\author{\firstname{M.~V.} \surname{Popov}}
\email{mpopov@asc.rssi.ru}%
\affiliation{%
\mbox{Astro Space Center of the Lebedev Physical Institute, Profsoyuznaya 84/32,} Moscow, 117997 Russia
}%

\author{\firstname{V.~A.} \surname{Soglasnov}}
\affiliation{%
\mbox{Astro Space Center of the Lebedev Physical Institute, Profsoyuznaya 84/32,} Moscow, 117997 Russia
}%

\author{\firstname{Y.~Y.} \surname{Kovalev}}
\affiliation{%
\mbox{Astro Space Center of the Lebedev Physical Institute, Profsoyuznaya 84/32,} Moscow, 117997 Russia
}%
\affiliation{%
\mbox{National Radio Astronomy Observatory, P.O. Box 2, Green Bank, WV 24944, USA}
}%
\affiliation{%
\mbox{Max-Planck-Institut f\"ur Radioastronomie, Auf dem H\"ugel 69, 53121 Bonn, Germany}
}%

\author{\firstname{N.} \surname{Bartel}}
\affiliation{%
\mbox{York University, Department of Physics and Astronomy, 4700 Keele Street, Toronto,} Ontario M3J 1P3 Canada
}%

\author{\firstname{W.} \surname{Cannon}}
\affiliation{%
\mbox{York University, Department of Physics and Astronomy, 4700 Keele Street, Toronto,} Ontario M3J 1P3 Canada
}%

\author{\firstname{A.~Yu.} \surname{Novikov}}
\affiliation{%
\mbox{York University, Department of Physics and Astronomy, 4700 Keele Street, Toronto,} Ontario M3J 1P3 Canada
}%

\begin{abstract}
VLBI observations of the Crab pulsar with the 64-m radio telescope at Kalyazin (Russia) 
and the 46-m radio telescope of the Algonquin Radio Observatory (Canada) at 2.2~GHz and single-dish 
observations of the
millisecond pulsar B1937+21 with the GBT (USA) at 2.1~GHz were conducted to probe the interstellar
medium and study the properties of giant pulses.
The VLBI data were processed with a dedicated software correlator, which allowed us to obtain the 
visibility of single giant pulses. 
Two frequency scales of 50 and 450~kHz were found in the diffraction spectra of giant pulses from
the Crab pulsar. The location of the scattering region was estimated to be
close to the outer edge of the nebula. No correlation was found between the power spectra of giant pulses at left-
and right-hand circular polarization. We explain this lack of correlation through the influence of the strong magnetic field 
on circularly polarized emission in the region close to the Crab pulsar.

Combining the measurement of the decorrelation bandwidth with that of the scattering time of giant
pulses for B1937+21, we found three frequency scales of 1.7, 3.8, and 16.5~MHz.
The scattering time of giant pulses of B1937+21 at 2.1~GHz was found to be $40\pm 4$~ns.
We obtained an upper limit of the intrinsic width of giant pulses from B1937+21 of less than 8~ns.
The frequency dependences of the scattering times for the Crab pulsar and PSR~B1937+21 were found
to be different. They are characterized by exponents of $-3.5$ and $-4.2$, respectively. We attribute the
difference to the large influence of scattering in the Crab nebula.
\end{abstract}

\maketitle

\section{Introduction}
Pulsar radio emission on its way to Earth is affected by the turbulent
interstellar medium (ISM) which is observed through  
dispersion, pulse and angular broadening, intensity scintillations, diffraction in
radio spectra, and other effects. Given these effects it is very difficult
to study the original properties of the unaffected emission. But, on the
other hand, they are very useful to probe the ISM itself. The most encouraging
probes to study the ISM seem to be giant radio pulses (GPs) due to their huge flux densities
(up to tens of MJy), short durations (of order nanoseconds), and very high
degree of polarization. Below we are studying the properties of the ISM with
the GPs from the millisecond pulsar B1937+21 and the Crab pulsar.
We have also used the influence of the ISM on regular pulsar emission
to determine an upper limit on the intrinsic width of GPs.

\section{Observations}

In this paper we present a preliminary analysis of a) 
VLBI observations of the Crab pulsar with the 64-m radio telescope at Kalyazin (Russia)
and the 46-m radio telescope of the Algonquin Radio Observatory (ARO, Canada) at 2244~MHz
and b) single-dish observations of B1937+21 with the Robert C. Byrd Green Bank Telescope (GBT, USA) at 2.1~GHz.

VLBI observations of the Crab pulsar were conducted on July 19--20, 2005
with the S2 VLBI system~\cite{cannon1997} providing continuous recording in two 16-MHz bands.
We recorded upper and lower sidebands (USB and LSB) at the central frequency of 2244~MHz
at ARO where only one polarization channel (left circular) was available, and recorded
only USB for both left- and right-hand circular polarization (LCP, RCP) at Kalyazin.
Such configuration was used in order to overcome possible inconsistencies
in the designation of polarization channels at the two observatories.
Two observing sessions were carried out,
each session lasting for about 3 hours  with additional 
10 minutes each at the start and the end of the session for observations of the 
continuum calibration source DA193.

Observations of the millisecond pulsar B1937+21 were done on June 7, 2005 
with the GBT at a frequency of 2.1~GHz
in both LCP and RCP with a time
resolution of 8\,ns.  Four adjacent 16-MHz channels (2052--2116 MHz)
at each polarization were digitized simultaneously with 2-bit
sampling at the Nyquist rate. The Mark5A data acquisition system was
used for the first time in \emph{single-dish} observations with the
GBT~\cite{memo236}. This allowed us to obtain continuous and uniform recording for about
7.5~hours with a data rate of 512\,Mbps. The quasar 3C286, the radio
source 3C399.1, and the planetary nebula NGC~7027 were observed as
well, for flux density and polarization calibration.  The system
temperature in all 8 separate frequency channels was about 23~K.


\section{VLBI observations of giant pulses from the Crab pulsar}
\subsection{Peculiarities of the data reduction}
In our observations we expected to find large sudden amplitude increases 
of the cross-correlation function (CCF) at short and random time intervals when
GPs occurred. GPs have typical time scales of a few microseconds, and their
peak flux densities can exceed a million Jy after
dispersion removal. To provide
dedispersion and allow for short integration we developed dedicated software
for the following functions: time alignment of recorded signals,
decoding of the 2-bit sampling with corrections for current bit statistics,
predetection dedispersion, fringe compensation, calculation of the CCF via FFT,
and postcorrelation analysis. The software is based on
a geometric delay model, which we used in a form of the polynomial
\begin{equation}
\label{obs/0}
\tau_\mathrm{g}(t)=a+b(t-t_0)+c(t-t_0)^2+d(t-t_0)^3~.
\end{equation}
In our simple case of two radio telescopes, $\tau_\mathrm{g}(t)$ represents the
relative time delay with the time at the Kalyazin radio telescope used as a reference.
The coefficients of the polynomial were calculated with the
software ``WinEra'', provided by the Institute of Applied Astronomy in St.~Petersburg~\cite{kras_n_vas1997}.
To obtain the fringe frequency we multiplied the signal 
recorded at ARO with the complex function 
$\exp(-j2\pi\nu_0\tau_\mathrm{g}(t))$. In our case the residual phase delay can be
written as
\begin{equation}
\label{obs_2}                                                 
\phi_{12}(t)=2\pi[\nu_0\Delta\tau_\mathrm{g}(t)+\Delta\nu_0t+\phi_\mathrm{V}(t)]~,
\end{equation}
where $\Delta\tau_\mathrm{g}(t)$ is the residual delay due to an inaccurate
knowledge of the source or radio telescope position; $\Delta\nu_0$ the
difference between the LO-frequencies at the observatories; and $\phi_\mathrm{V}(t)$ 
the phase delay produced by the atmosphere, ionosphere, and scattering
medium.

The data reduction was conducted at the Astro Space Center (Moscow, Russia).
The data from the S2 video tapes were
copied to hard disk using the Radioastron Data Recorder 
Interface (RDR). The time delay compensation was implemented in two
stages removing first integer and then fractional delays, the latter
through phase correction of the spectrum together with dispersion
removal. The dispersion smearing, $\tau_\mathrm{DM}$, over a 16-MHz
band at a frequency of 2244~MHz for the used value of
$\mathrm{DM}=56.737$~pc cm${}^{-3}$~\cite{jodrell} is equal to $660~\mu$s. The predetection
dispersion removal technique~\cite{hankins} requires that the time interval, $T$, used for
compensation must be longer than  $\tau_\mathrm{DM}$.
We used $T=8192~\mu$s (number of samples: N=262,144) for dedispersion and fringe compensation.
The CCFs were computed via FFT (FX-correlator) on time arrays,
$\Delta T$, of 32~$\mu$s duration (N=1024).
The normalized CCF magnitude was derived as
$M_\mathrm{ccf}=(\sqrt{R_\mathrm{ccf}^2+I_\mathrm{ccf}^2})/(\sqrt{\mathrm{acf}_1\cdot \mathrm{acf}_2})$,
where $R_\mathrm{ccf}$ and $I_\mathrm{ccf}$ are, respectively, the real and imaginary components of the CCF,
and $\mathrm{acf}_1$, $\mathrm{acf}_2$ are the values of the autocorrelation functions (ACF) at zero-lag for
the two observing sites for the given time interval $\Delta T$.
The expected value of the rms deviation of CCF magnitudes
can be estimated as  $\sigma_\mathrm{CCF}=1/\sqrt \mathrm{N}=0.031$. In our
search for CCFs on GPs we used the threshold of $6\sigma=0.186$ in
a restricted range of time lags of $\pm 0.1 \mu$s,
and CCFs with magnitudes greater than the threshold were selected
for postcorrelation analysis.
The visibility amplitude was corrected for the signal-to-noise ratio (SNR) by multiplying
it with the factor $R=\frac{\sigma_\mathrm{t1}\sigma_\mathrm{t2}}
{\sqrt{((\sigma^2_\mathrm{t1}-\sigma^2_\mathrm{off1})(\sigma^2_\mathrm{t2}-\sigma^2_\mathrm{off2}))}},$
where $\sigma_\mathrm{t1}$ and $\sigma_\mathrm{t2}$ are GP on-pulse rms deviations for a given $\Delta T$
and $\sigma_\mathrm{off1}$ and $\sigma_\mathrm{off2}$ are the equivalent values
for the noise portion of the records. The visibility phase was determined as
$\phi_\mathrm{V}= \arctg (I_\mathrm{ccf}/R_\mathrm{ccf})$ with $I_\mathrm{ccf}$ and $R_\mathrm{ccf}$ taken at the 
maximum magnitude of the CCF.

\subsection{Scintillations of the visibility function}
\label{scintV}

The scattering of the initially coherent electromagnetic radiation
from a pulsar caused by the inhomogenities of the cosmic plasma produces
angular broadening  of the source image,
time smearing  of individual
pulses, and diffraction distortion  of the radio spectrum. A simple
model of a thin screen located midway between Earth and the
pulsar provides the following relations~\cite[][p.~91]{lorimer_n_kramer}:
$\tau_\mathrm{sc}=\theta_\mathrm{sc}^2D/4c$, and $2\pi\tau_\mathrm{sc}\Delta f=1$,
where $\tau_\mathrm{sc}$ is the time smearing parameter, $\theta_\mathrm{sc}$ the angular diameter of the scattered image,
$D$ the distance to the pulsar, and $\Delta f$ the decorrelation
bandwidth. The maximum resolution of our two-element interferometer
was equal to $\theta_I=\lambda/\mathrm{B}=4$~mas (B=6750 km),
while the angular diameter of the
scattered image corresponding to $\tau_\mathrm{sc}=200$~ns
is estimated to be about 0.4~mas.
Therefore, we did not expect to readily resolve 
the scattered image, but we hoped to relate certain effects
of the scattering to the time and frequency behavior of the visibilities
measured for the GPs.

First, the magnitudes of the visibilities for all strong GPs
($\mathrm{SNR}_\mathrm{CCF}>10$), when corrected for finite SNR (factor $R$ above),
are close to $0.92\pm 0.05$. So, we did not detect any clear decorrelation
caused by the diffraction pattern produced in the plane of observations
by interstellar scattering. In fact, our observations
are similar to "snapshot" observations, since we obtain essentially instant
values of the visibility (formal integration time is $32~\mu$s).
However, in the frequency domain, the integration was done over a total bandwidth 
of 16-MHz, covering many diffraction scintles in the spectrum. We will see below
that at least two scales of diffraction scintles
were found in the radio spectra with widths of about 50 and 450~kHz.
Following Cordes et~al.~\cite{cordes_etal2004} one will find about
$0.4B_\mathrm{tot}/\Delta f$ scintles across a total band, $B_\mathrm{tot}$. In our case
the numbers are about 130 and 15 scintles for narrow and broad
structures, respectively. Therefore, we are averaging in fact complex
vectors of visibility over that number of scintles, and still have
them coherent (no loss of correlation magnitude). This result
is quite natural, since our two-element interferometer did not
resolve the scattered disk.

\begin{figure}[htb]
\setcaptionmargin{5mm}
\includegraphics[scale=0.67,angle=270]{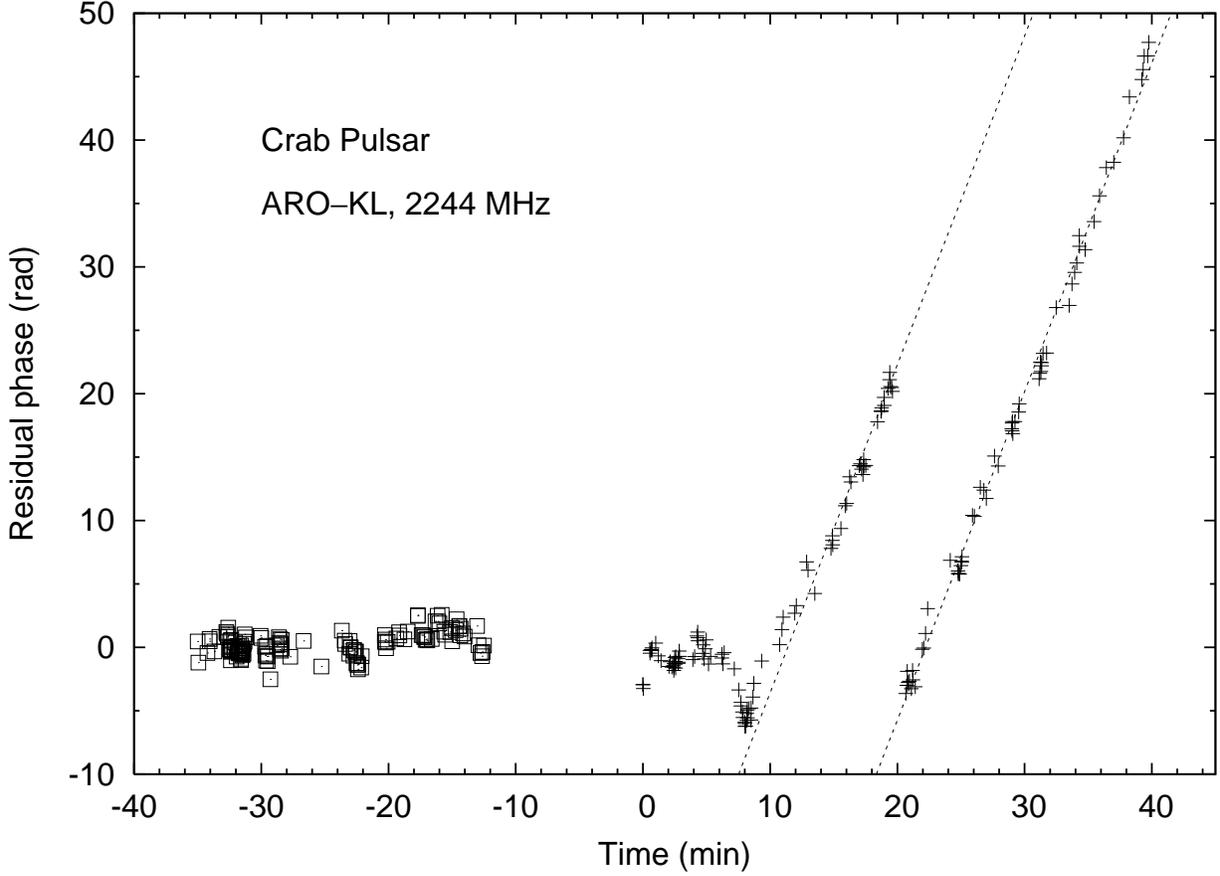}
\caption{
The residual phase of visibility in a time interval of about an hour during
VLBI observations of the Crab pulsar between ARO and Kalyazin (KL) at 2244~MHz.
Squares and pluses represent the residual phase in radians before and after
the gap caused by the poiting problems at Kalyazin. Dashed lines represent
the linear drift in phase at the rate of $0.042$~rad/s due to the lock-on of LO at
one of the observatories.
}
\label{visib_time}
\end{figure}

Let us now consider the time behavior of the residual phase of the visibility
presented in Figure~\ref{visib_time} for the period of about an hour.
At the beginning of the observing session one can see a constant residual phase with only relatively
small random variations
relative to the zero level with an rms deviation $\sigma\sim 0.5$~rad.
This behavior continues during several minutes after an observing gap
caused by troubleshooting of pointing at Kalyazin. Then there occurs a
negative jump in phase followed by a linear drift at the rate
of 0.042 rad/s. The reason of the drift may be the lock-on of the
local oscillator system at one observing site with a relative
inaccuracy of $3\times 10^{-12}$.

\begin{figure}[htb]
\setcaptionmargin{5mm}
\includegraphics[scale=0.47]{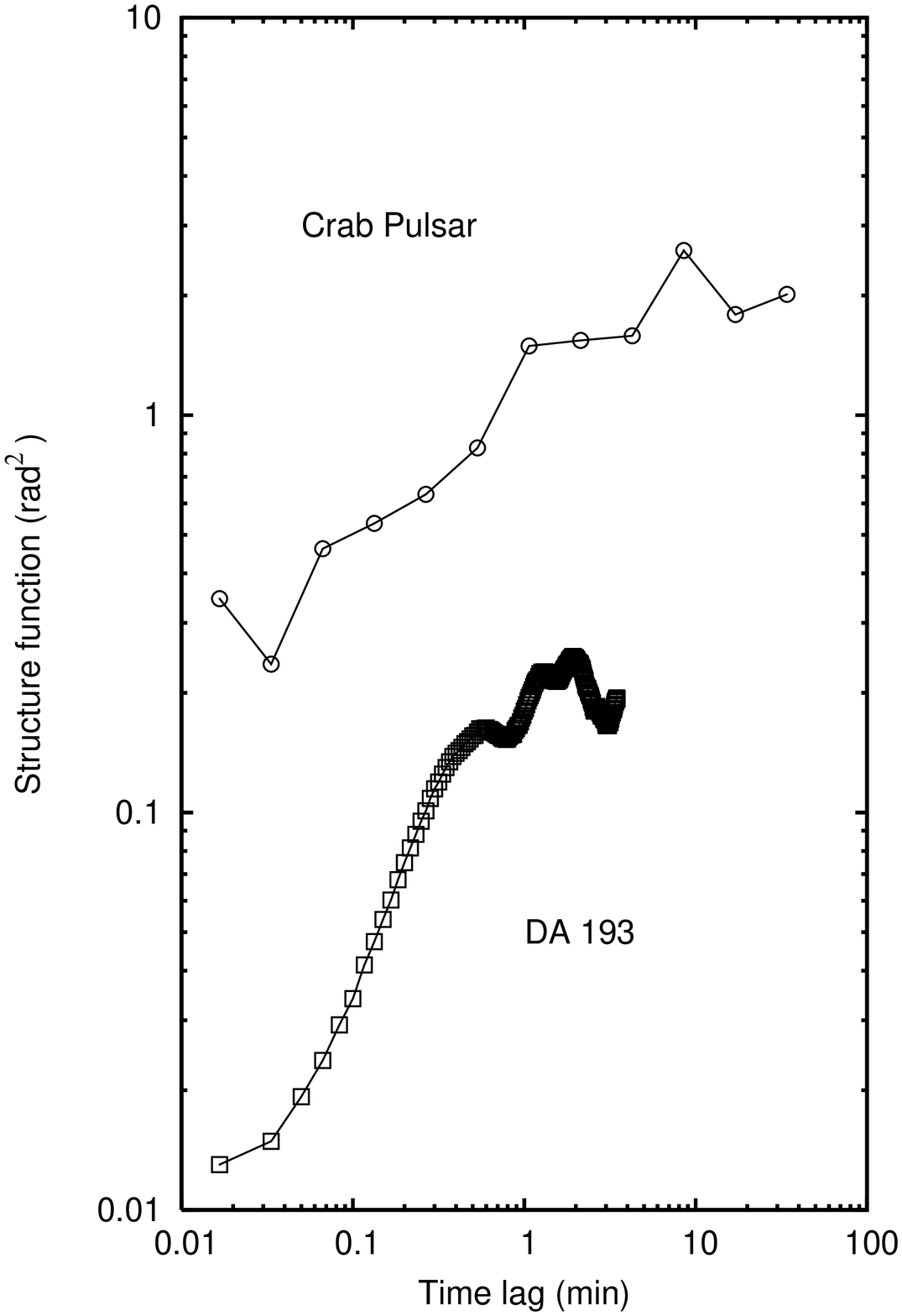}\includegraphics[scale=0.47]{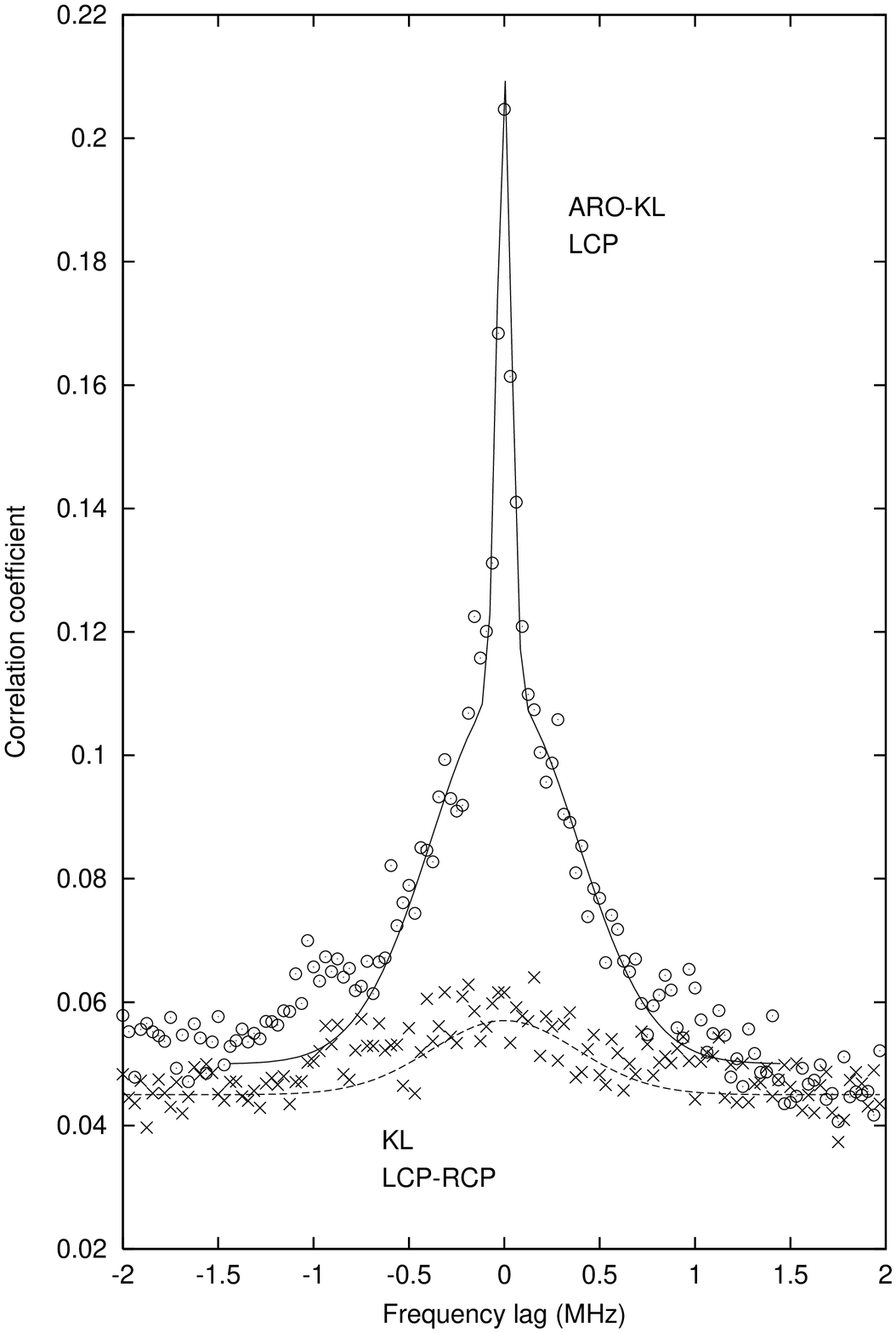}
\caption{
{\bf Left.} Phase structure function of visibility phase variations 
for the Crab pulsar (circles) and quasar DA193 (squares). {\bf Right.}
The average CCF between power spectra of GPs recorded at LCP at Kalyazin
and ARO (circles), and the average CCF between power spectra of the
same GPs recorded at Kalyazin at LCP and RCP (crosses). The solid line is the approximation
of the average CCF with two gaussians having half-widths of 50 and 450~kHz.
The dashed line indicates the broad feature of 450~kHz marginally seen on the
average CCF between power spectra of GPs recorded at Kalyazin at LCP and RCP.
The frequency resolution is 30~kHz.
}
\label{struct}
\end{figure}

The phase structure function
$D_{\phi}(\tau)=\langle[\phi_\mathrm{V}(t)-\phi_\mathrm{V}(t+\tau)]^2\rangle_\tau$
for the Crab pulsar
is shown in Figure~\ref{struct} (left)
together with the phase structure function obtained
for the continuum source DA193 observed
at the beginning of the session. One can see that the
phase structure functions for the pulsar and the contimuum source
are drastically different.
The residual rms phase variations of the visibility function for
DA193 are about 0.1 rad for short time intervals ($\tau < 1$~c)
between measurements, and they are saturated at long time intervals ($\tau > 1$~min)
at a level of about 0.4~rad in rms. We attribute these variations to
the effects of the atmosphere and ionosphere.
The residual rms phase variations
of the visibility function for the Crab pulsar are about 5 times larger ($\approx 0.5$~rad)
than those caused by the atmosphere and the ionosphere, even for the shortest time intervals
(there were 73~ GPs separated by less than
1~second in time).
Since we do not apply any time averaging
of visibility, the observed phase variations can be
interpreted as random angular displacements of the point source in the range
restricted by the scattering disk with diameter $\theta_\mathrm{sc}$.
Using relation~(\ref{obs_2}) one can convert the phase variations to
geometrical delay variations via $\Delta \tau_\mathrm{g}=\Delta\phi_\mathrm{V}/2\pi\nu_0$,
thus obtaining $\Delta\tau_\mathrm{g}=3.5\times 10^{-11}$~c. On the other hand,
we can convert such variations in $\tau_\mathrm{g}$ to a corresponding
scattering diameter $\Delta\theta_{\tau}$ via the relation~\cite[][p.~376]{tompson_mor_swen}
$$\frac{\Delta\tau_\mathrm{g}}{\Delta\theta_{\tau}}=\frac{\mathrm{B}}{c}\cos\theta~,$$
where $\theta$ is the angle between the source vector and the
plane perpendicular to the baseline. For the part of the observing session
under consideration, $\theta$ was equal to $15^\circ$,
resulting in a projected baseline length of $\mathrm{B}_\mathrm{p}=\mathrm{B}\cos\theta=6500$~km.
Finally, we got an estimate for the scattering diameter of
$\Delta\theta_\tau=(c\Delta\tau_\mathrm{g})/\mathrm{B}_\mathrm{p}=1.5\times10^{-9}$~rad~$=0.3$~mas.
We would like to emphasize that the technique of "phase self referencing"
used above provides at least a tenfold better sensitivity of estimating the scattering diameter than the formal
resolving power of our two-element interferometer.

With the $\Delta\theta_\tau$ estimate in hand we can get additional
information on the location of the scattering region (screen).
Following the technique used by \cite{desai1992} in their analysis
of the specle hologram of the scattering material along the line of sight
to the Vela pulsar, we get
$$\frac{dD}{D-d}=\frac{2.64\times10^{11}}{\Delta f(\Delta\theta_\tau)^2}~.$$
Here $D$ is the distance to the pulsar ($D=2$~kpc), $d$ 
the distance from the observer to the scattering screen,
and $\Delta f$ the decorrelation bandwidth. We will show below
that the radio spectrum of GPs indicates two scales
of diffraction pattern with $\Delta f_\mathrm{n}=50\pm10$~kHz (narrow feature),
and $\Delta f_\mathrm{b}=450\pm10$~kHz (broad feature), with both structures having
approximately equal modulation indices (see Figure~\ref{struct}, right).
It appears reasonable to suggest that the major part of the
detected angular
broadening is related to the narrow diffraction pattern while the
broad features in the spectra cause angular
deflections on  smaller scales.
In this case the distance from the pulsar to the screen
constitutes $2.5\times 10^{-3}$ of $D$, or about $5$~pc
($\approx 3$~radii of the Crab nebula).

Now we will give some explanations on the technique of the
measurement of the decorrelation bandwidth in the spectra
of GPs. Since the intrinsic duration of GPs ($\tau_\mathrm{GP}$),
after dispersion removal, is only a few
microseconds, the frequency resolution in the radio
spectra of GPs is restricted by the factor $1/\tau_\mathrm{GP}\approx 0.3-3$~MHz.
Prior to dedispersion, the smearing time over a 16-MHz band is equal to $660~\mu$s
giving a frequency resolution of 1.5~kHz but with low SNR.
We chose to remove the dispersion only partly, corresponding to 
$80\%$ of the DM value.
This procedure leaves time smearing of $128~\mu$s giving a frequency resolution of
7.8~kHz. With this approach power spectra were computed for all
GPs with $\mathrm{SNR} > 7\sigma$ ($107$ in total). 
Figure~\ref{struct} (right) shows two different CCFs: the average CCF between power 
spectra of GPs recorded at LCP at Kalyazin and ARO, and the average CCF
between power spectra of the same GPs recorded at Kalyazin at LCP and RCP. 
The approximation of these average CCFs with two gaussians gives 
values for the decorrelation bandwidths of $50$ and $450$~kHz.
To estimate the characteristic time scale of scintillations we
computed the average CCF between power spectra separated in time
by a certain time lag. The narrow band feature in the CCF was only marginally
detected at a level of about $2\sigma$ even for the closest GPs with a time
separation less than 1 second.

Most interesting in Figure~\ref{struct} (right) is the absence of any
correlation between diffraction spectra of GP emission in LCP and RCP. 
In contrast, such correlation was found for
B1937+21 (see Section~\ref{DecBand}).
In the presence of a magnetic field the refractive indices of LCP and RCP waves are 
different. In principle, this can lead to different scattering for radio emission
at LCP and RCP. However, the magnetic field 
strengths in the ISM ($10^{-5}$--$10^{-6}$~G)
and in the Crab nebula ($10^{-3}$~G) are too small to affect the 
scattering. We believe that the absence of correlation for the Crab pulsar
could be caused in the region near the Crab pulsar 
where the low-frequency 33-Hz emission is converting to the pulsar wind.
Just outside the light cylinder the magnetic field has a considerable tangential 
component of $\sim 10^6$~G, so the Larmour frequency, $\nu_\mathrm{H}$, is much larger 
than our observing frequency, $\nu$,
of 2244~MHz and remains larger up to distances of 30--50 light cylinder radii.
The magnetic field does not have an effect on a linearly polarized wave of which the
E-vector is parallel to the external H vector (ordinary wave).
The refractive index is defined by the usual expression for a cold plasma. 
A wave with E-vector orthogonal to the external H-vector  is an extraordinary wave.
If $\nu_\mathrm{H}\gg\nu$, then its refractive index is very close to $1$,
and the influence of the tangential magnetic
field compensates almost completely the normal dispersion. The intensity of a
circularly polarized wave has a periodic modulation with frequency.
The intensity maxima of LCP coincide with the intensity minima of RCP and vice 
versa. This is likely the reason why
the correlation between LCP and RCP scintillation fringes in the frequency 
domain is strongly suppressed at zero-lag but may rise at non-zero lag 
($\sim 1$~MHz).
At a given frequency, $\nu$, the period depends only on the column density number of 
free thermal electrons $N_\mathrm{e}$: 
$P_{\nu}\approx 220 D\nu^2/N_\mathrm{e}$. 
Since $N_\mathrm{e} \sim 4 \cdot 10^{15}$~cm${}^{-3}$, it corresponds to a dispersion 
measure of $\sim 1.3 \cdot 10^{-3}$~pc cm${}^{-3}$. This value is approximately equal to 
the range of variations of the dispersion measure observed for the Crab pulsar~\cite{isaac1977}.


\section{Measurements of pulse scattering of the PSR B1937+21}
\subsection{Data processing}

Data processing was performed both at the Astro Space Center and
York University (Toronto, Canada). Just after the observations, the
raw data were split into separate pieces of $10^9$ bytes and copied from the Mark5 '8-pack' 
disk modules to external 1-TB disks. Then these disks were shipped to the home institutions, and the
Mark5 data were decoded and the obtained signal was corrected for instantaneous
bit statistics and amplitude bandpass irregularities and then coherently
dedispersed. After dedispersion we were searching for giant pulses in every topocentric
period with the detection threshold of $17\sigma$ in every 16-MHz band.
The processing procedure and the detection criteria are described in 
detail in~\cite{kondratiev2006}.

In 5.5 hours of pulsar data processed to date we found 6334 giant pulses stronger 
than 205 Jy in the 16-MHz bands. For further processing with the goal of studying GP scattering
the 22 strongest GPs were selected which reached a peak flux density of $> 1.2$~kJy in the
total 64-MHz band.

\subsection{Decorrelation bandwidth and pulse broadening}
\label{DecBand}

Even observing at the relatively high frequency of 2100~MHz, the interstellar 
scattering of pulsar radio emission is still notable, especially when
the signal is recorded with as high a time resolution as in our case.
The scattering time, $\tau_\mathrm{sc}$, at our frequency may be estimated by extrapolating the
scattering time measured at 1650~MHz~\cite{soglasnov2004}.
Assuming $\tau_\mathrm{sc}\sim\nu^{-4.4}$ (see Section~\ref{freqdep}) we 
obtain $\tau_\mathrm{sc}\approx 40$~ns at our frequency or 5 samples for the
total 64-MHz band. This value corresponds to a characteristic decorrelation
bandwidth of $\sim4$~MHz. First, to make accurate measurements of the decorrelation
bandwidth and scattering time, we constructed the average ACFs and CCFs  for the power spectra in LCP 
and RCP for the 22 selected strong GPs~\cite[see][]{kondratiev2006}. It should be 
mentioned, that in contrast to the VLBI observations of GPs
from the Crab pulsar (see Section~\ref{scintV}) the correlation between spectra
in LCP and RCP for GPs from the B1937+21 is significant.
Not only one, but two frequency scales of $3.79\pm 0.04$~MHz and
$16.5\pm 0.8$~MHz were found~\cite{kondratiev2006}.

\begin{figure}[htb]
\setcaptionmargin{5mm}
\includegraphics[scale=0.67,angle=270]{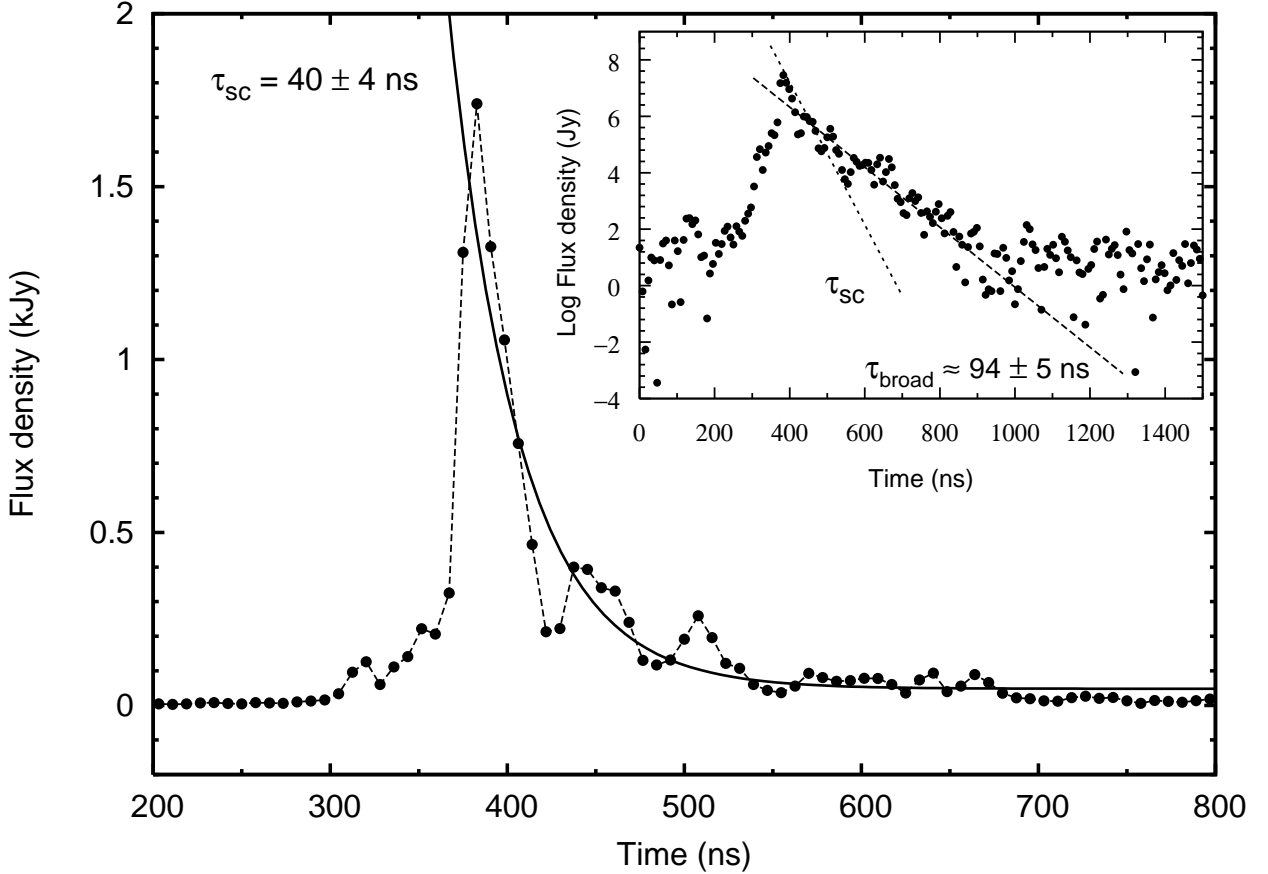}
\caption{
The average profile of 20 strong GPs of B1937+21 based on observations with the GBT at 2.1~GHz (dashed line).
Solid line shows the least-square fit to the exponential scattering tail of the profile with an estimated
scattering time $\tau_\mathrm{sc} = 40\pm 4$~ns. The inset plot represents the same
GP average profile (points) in a larger time window and with the Y-axis drawn in logarithmic scale.
The short-dashed line represents the fit from the main plot, and the long-dashed line is a least-square fit 
to the profile tail in logarithmic scale with a measured broad-scale scattering 
time $\tau_\mathrm{broad}\approx 94\pm 5$~ns.
Both main and inset plots are plotted with the sampling interval of 7.8125~ns.
}
\label{exptail}
\end{figure}

To confirm the estimated scattering time of 40 ns, which is also derived
from the measured decorrelation bandwidth of 3.8~MHz, we can also analyse directly
GP shapes and measure their scattering time. To avoid random noise
fluctuations of individual GPs, we fold together 20 GPs from the set we used
for the CCF analysis dropping two GPs with relatively complex shapes that
could be intrinsic to the pulses themselves rather than induced by the scattering in the ISM.
This average profile of GPs is shown in Fig.~\ref{exptail}.
Fitting an exponential curve to the profile tail we obtained for the exponent
$\tau_\mathrm{sc} = 40\pm 4$~ns which is the scattering time we could now assume for our observations
at our observing frequency. 
It is evident that, when presenting
the pulse profile with the flux density axis drawn in a logarithmic scale,
the exponential tail appears as a straight line.
This line is clearly seen in the inset plot in Fig.~\ref{exptail}.
The measured slope gave us the broad-scale scattering time of $\tau_\mathrm{broad}\approx 94\pm 5$~ns 
that gives a corresponding
decorrelation bandwidth of $1.7\pm 0.1$~MHz.

Thus, analysing the power spectra and the average profile of GPs from B1937+21, we found
three different frequency scales of 1.7, 3.8, and 16.5~MHz.
It is likely that these scales correspond to different screens of enhaned interstellar plasma 
along the line of sight between the pulsar and Earth (spiral arms and the local screen).

\subsection{Superresolution of giant pulses of the PSR B1937+21}

As was concluded above, the width of giant pulses is affected
by scattering. However, it turns out that in our particular case
the ISM itself helps to put constraints on the intrinsic
width of GPs.
It is obvious that both regular
and GPs travel along the same path and probe the same ISM.
Thus, we can use
the regular emission to correct the GP emission for the influence of the ISM.
The apparent shape of the GPs is a result of
convolution of the intrinsic GP waveform
with the transfer function of the interstellar medium.
We can use the diffraction spectrum obtained for regular emission
as the transfer function to determine the intrinsic
width of GPs by calculating the response on the infinitely
short $\delta$-pulse passed through a 
scattering medium. 

\begin{figure}[htb]
\setcaptionmargin{5mm}
\includegraphics[scale=0.67,angle=270]{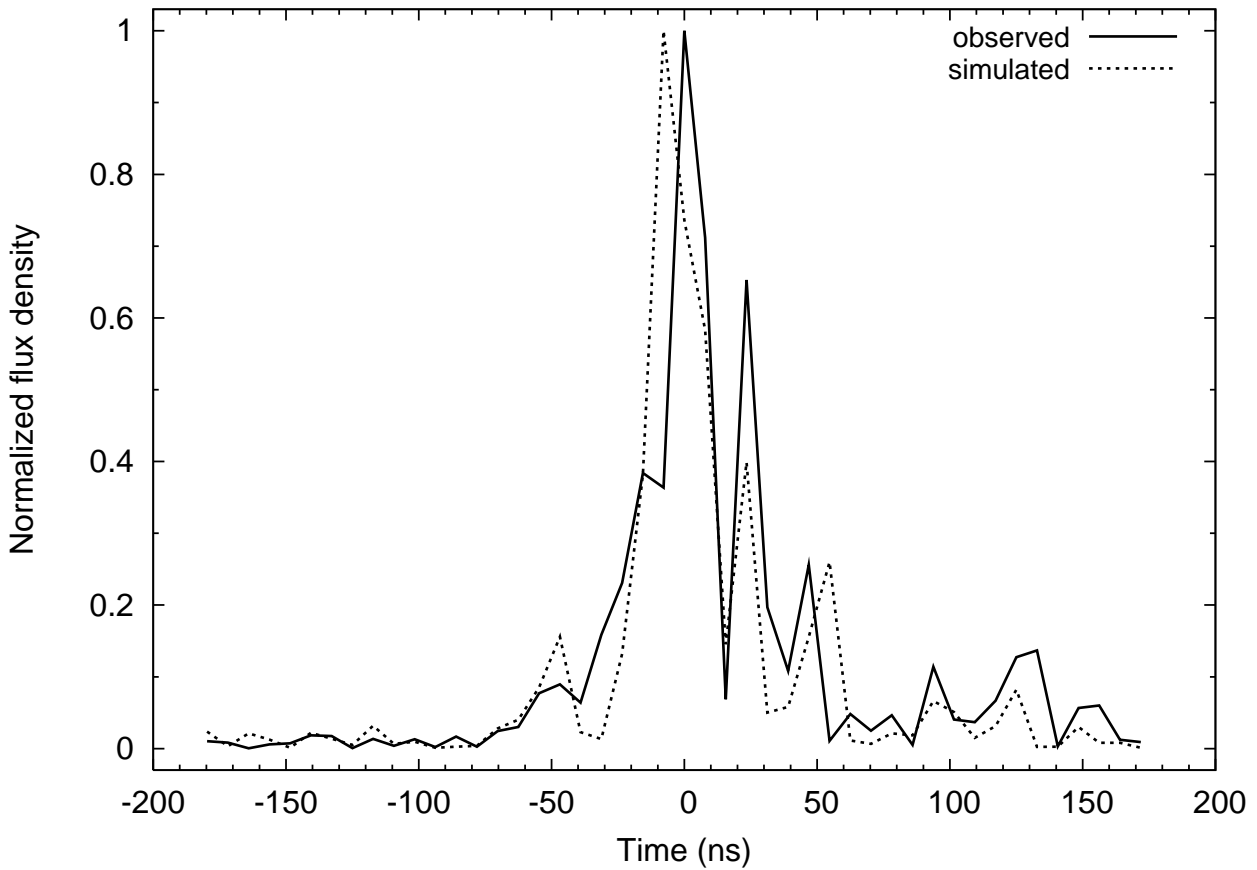}
\caption{
The strongest GP with a peak flux density of 10~kJy in the total 64-MHz band
from B1937+21 recorded with the GBT at 2.1~GHz (solid line). The dashed line represents the simulated
response from a $\delta$-pulse passed through the ISM using the diffraction spectrum of regular
emission as its transfer function. The observed and simulated pulses are shown with the
sampling interval of 7.8125~ns.
}
\label{superres}
\end{figure}

Fortunately, the amplitude and the phase of the transfer function
are not independent but related via the Hilbert transform~\cite[][p.~553]{gonor1977}. 
It allows us to obtain the waveform of the initially short but scattered
pulse directly from the diffraction spectrum of the regular emission.
In Fig.~\ref{superres} the strongest  GP is shown together with the simulated response 
from a $\delta$-pulse. The comparison between the observed and simulated pulse shows very good agreement in the
number of peaks, their relative amplitudes, and their positions.
Such a good agreement is a strong evidence that the intrinsic width of GPs
is less than our sampling interval of 8~ns. 

This measured upper limit on the width of GPs allows us to put an upper limit on the size
of the GP emitters in case GPs are temporal fluctuations. Then the size, d, of the giant
pulse emitter, is $d < c\tau$, or less than 2 meters if relativistic effects can be ignored.
Such a short duration together with the high peak flux density of 10~kJy gives a
brightness temperature of the strongest GP of greater than $10^{37}$~K.

\subsection{Frequency dependence of scattering time}
\label{freqdep}

\begin{figure}[htb]
\setcaptionmargin{5mm}
\includegraphics[scale=0.67,angle=270]{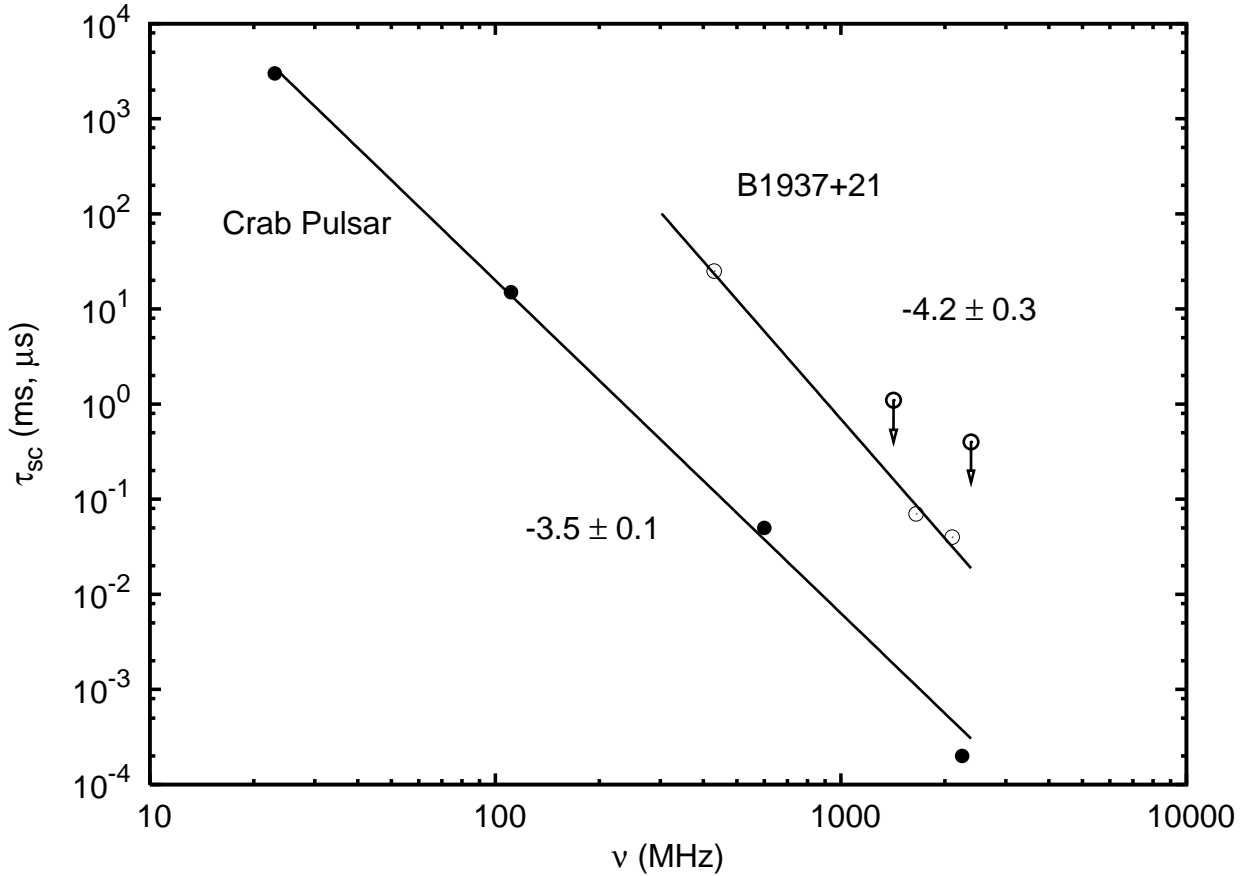}
\caption{
The frequency dependence of the scattering time, $\tau_\mathrm{sc}$, for B1937+21 (in~$\mu$s) and 
the Crab pulsar (in~ms). The data points are from this paper and 
\cite{popov2006a, cognard1996, kink2000, soglasnov2004} (for more detail, see text).
The solid lines represent least-squares power-law fits with indices of $-4.2\pm 0.3$ for B1937+21
and $-3.5\pm 0.1$ for the Crab pulsar.
}
\label{taunu}
\end{figure}

Having measured the scattering parameters 
allows us to test the form of the spectrum of 
inhomogeneties of the electron density in the ISM. To do so we have to also use
the measurements of the scattering time (or decorrelation bandwidth) at other frequencies.
For B1937+21 we measured the scattering time to be 65~ns at 1650~MHz~\cite{soglasnov2004}.
Cognard et~al.~\cite{cognard1996} obtained the scattering time of $25~\mu$s at 430~MHz with
observations at Arecibo. Kinkhabwala \& Thorsett~\cite{kink2000} put an upper limit for
the scattering time of GPs at 1420 and 2380 MHz to be 1.1 and $0.4~\mu$s, respectively.
Fitting a power-law to these values results in an 
index of $-4.2\pm 0.3$ (see Fig.~\ref{taunu}).
This value for the index correponds very well to the Kolmogorov spectrum of electron density.

For the Crab pulsar we used the values of scattering time obtained during the simultaneous
observations of GPs at the frequencies of 23, 111, and 600~MHz~\cite{popov2006a}
together with the measurement of the scattering time at 2244~MHz from our VLBI observation (see Section~\ref{scintV}).
These values of scattering time are: 3~s at 23~MHz, 15~ms at 111~MHz, $50~\mu$s at 600~MHz, and
$0.2~\mu$s at 2244~MHz. These values together with a line representing the best fit are shown in Fig.~\ref{taunu}.
The measured power-law index is equal to $-3.5\pm 0.1$ which is much lower than the value of $-4.4$ for the
Kolmogorov spectrum, and even lower than $-4$ for the Gaussian spectrum. This lower value for the Crab pulsar
may be due to the huge extra scattering in the surrounding
Crab nebula.

\section{Conclusions}

Different frequency dependences of $\tau_\mathrm{sc}$ were found for
the millisecond pulsar and for the Crab pulsar with 
exponents of $-4.2\pm0.3$ and $-3.5\pm0.1$, respectively. We attribute the
difference to the notable influence of scattering in the Crab
nebula. The location of the scattering region was estimated to be
close to the outer edge of the Crab nebula. The estimate is based on an analysis
of the scattering diameter ($\theta_\mathrm{sc}\approx 0.3 $~mas),
obtained through  VLBI phase self-reference measurements, and the
scattering time, determined from the diffraction pattern in radio
spectra of GPs ($\Delta f = 50$~kHz). A significant difference
was found between the diffraction pattern in radio spectra of GPs
from the Crab pulsar emitted 
in RCP and LCP polarization channels. The effect is likely caused by the
propagation of radio emission through the magnetized plasma of the Crab
nebula. A simulation of the observed scattered waveform of GPs from
B1937+21 has shown that
the majority of GPs from this pulsar are shorter than 8~ns.


\begin{acknowledgements}
We thank Frank Ghigo for extraordinary help in the preparation of the observations with the GBT
and Konstantin Belousov and Andrey Chibisov for providing the operational S2-RDR play-back system.
The Robert C. Byrd Green Bank Telescope (GBT) is
operated by the National Radio Astronomy Observatory which is a
facility of the U.S. National Science Foundation operated under
cooperative agreement by Associated Universities, Inc. ARO is operated by the Geodetic Survey Division of
Natural Resources Canada.
VIK was a postdoctoral fellow at York University at the beginning of this project.  
Part of the project was done while YYK was a Jansky Fellow of the
National Radio Astronomy Observatory and a research fellow of the
Alexander von Humboldt Foundation.
This project was supported in part
by grants from the Canadian NSERC, the Russian Foundation for Basic
Research (project number 07-02-00074) and the Presidium of the Russian
Academy of Sciences ``Origin and evolution of stars and galaxies''.
\end{acknowledgements}


\end{document}